# Single-photon transport and mechanical NOON state generation in microcavity optomechanics

Xue-Xin Ren, Hao-Kun Li[#], Meng-Yuan Yan, Yong-Chun Liu, Yun-Feng Xiao[*], and Qihuang Gong

State Key Laboratory for Mesoscopic Physics and School of Physics, Peking University, Beijing 100871, P. R. China

**We investigate the single-photon transport in a single-mode optical fiber coupled to an optomechanical system in the single-photon strong-coupling regime. The single-photon transmission amplitude is analytically obtained with a real-space approach and the effects of thermal noises are studied via master-equation simulations. The results provide an explicit understanding of optomechanical interaction and offer a useful guide for manipulating single photons in optomechanical systems. Based on the theoretical framework, we further propose a scheme to generate the mechanical NOON states with arbitrary phonon numbers by measuring the sideband photons. The probability for generating the NOON state with five phonons is over 0.15.**

---

[#] Email address: hkli@pku.edu.cn
[*] Corresponding author: yfxiao@pku.edu.cn; URL: www.phy.pku.edu.cn/~yfxiao/index.html

# I. INTRODUCTION

The manipulation of light transport at the single-photon level is of central importance in both fundamental studies of quantum optics and many applied fields, such as quantum computation [1] and quantum information processing [2]. There have been numbers of studies probing single-photon transport properties in various interfaces, including cavity quantum electrodynamics [3-6] and surface plasmon [7, 8]. Benefiting from the strongly enhanced light-matter interaction, these interfaces provide practical architectures for controlling and manipulating single-photon transport.

Recently, a novel type of interaction between optical field and mechanical oscillators has attracted growing interests [9-15]. This type of light-matter interaction offers a promising platform for generating and manipulating quantum states of light and macroscopic oscillators. Experiments on quantum limited measurement [16], optomechanically induced transparency [17, 18] and mechanical ground state cooling [19, 20], represent remarkable steps toward investigations and applications of quantum optomechanics. With the rapid advancement of optical trapping and nanofabrication in recent years, optomechanical systems based on either ultracold atoms trapped in optical resonators [21, 22], superconducting circuits [23], or optomechanical crystals [24, 25] are approaching the single-photon strong-coupling regime [26-29], where the radiation pressure of a single photon significantly affects the movement of the mechanical oscillator. The strong interaction between optical field and mechanical oscillators could open up a new route to the implementation of single-photon transport. Thus, it is of importance to investigate the properties of single-photon transport in optomechanical systems.

In this paper, we investigate the single-photon transport in a single-mode optical fiber coupled to an optomechanical system which works in the regime of single-photon strong-coupling. In the stationary condition, the single-photon transmission amplitude is analytically solved. The results explicitly reveal the underlying physics of the photon transport. In contrast to a recent study [30], our formalism treats the problem in the real space and can obtain the output spectrums of any input photon spectral density distribution. Furthermore, using master-equation simulations, we study the effects of thermal noises on the single-photon transport.

Finally, we propose a scheme to generate the mechanical NOON states [31-33] via manipulating single-photon transport in optomechanical systems.

The paper is organized as follows. In Sec. II, we describe our model for single-photon transport. In Sec. III, we derive the analytical solution of photon transmission and discuss the physical process of the single-photon transport. The effects of thermal noises on the photon transport are presented in Sec. IV. In Sec. V, we depict a scheme for generating the mechanical NOON states and study the probabilities and fidelities of the scheme. Finally, our conclusions are shown in Sec. VI.

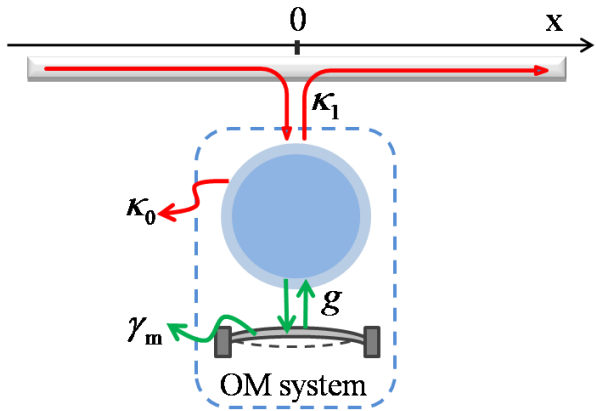

FIG. 1. Schematics of the coupling system. An optomechanical (OM) system is side-coupled to a single-mode optical fiber, in which photons propagate from left to right.

## II. MODEL

We study the transport of a single photon inside a single-mode fiber, which couples to an optomechanical system, as shown in Fig.1. We model the system with the Hamiltonian

$$H = H_{om} + H_{fib} + H_{om\text{-}fib}, \tag{1}$$

where $H_{om}$, $H_{fib}$, and $H_{om\text{-}fib}$ respectively describe the isolated optomechanical system, the fiber optical mode, and the coupling between them. The Hamiltonian of the isolated optomechanical system reads

$$H_{om} = \omega_c c^\dagger c + \omega_M b^\dagger b + g c^\dagger c (b + b^\dagger), \qquad (2)$$

where $c$ and $b$ are the annihilation operators for the cavity and mechanical mode, $g$ represents the single-photon coupling strength between the cavity and mechanical modes, $\omega_c$ denotes the optical resonance frequency of the empty cavity, and $\omega_M$ is the mechanical resonance frequency. In the real-space representation, by assuming a linear dispersion relation of the fiber optical mode, $H_{fib}$ and $H_{om-fib}$ can be expressed as

$$H_{fib} = \int_{-\infty}^{+\infty} dx (-i) v_g a^\dagger(x) \frac{\partial}{\partial x} a(x), \qquad (3)$$

$$H_{om-fib} = \int_{-\infty}^{+\infty} dx \delta(x) \sqrt{v_g \kappa_1} [a^\dagger(x) c + a(x) c^\dagger], \qquad (4)$$

where $a(x)$ represents the annihilation operator for a right-going photon at position $x$, $\kappa_1$ denotes the optical coupling rate between the cavity and the fiber modes, and $v_g$ is the group velocity of the photons in the fiber. The decoherence processes in the optomechanical system includes cavity and mechanical damping ($\kappa_0$ and $\gamma_M$), and environment thermal heating ($\Gamma_M = n_{th}\gamma_M$), where $n_{th}$ represents the mean thermal excitation phonon number.

### III. ANALYSIS OF SINGLE PHOTON TRANSOPRT

In this section, we provide an analytical solution of single-photon transmission amplitude in the regime $(g, \kappa_1) \gg (\kappa_0, \gamma_M, \Gamma_M)$, where the optomechanical decoherence processes are ignored. Initially, the mechanical oscillator is in number state $|m_0\rangle$, and a single photon with energy $\omega_0$ is input into the fiber. We investigate the photon transport in the stationary condition, described by the time-independent eigenequation

$$H|\omega_0, m_0\rangle = (\omega_0 + m_0 \omega_M)|\omega_0, m_0\rangle. \qquad (5)$$

Since the Hamiltonian $H$ conserves photon number, the interaction steady state $|\omega_0, m_0\rangle$ can be expanded in single-photon subspace, described as

$$|\omega_0, m_0\rangle = \sum_m \int_{-\infty}^{+\infty} dx f_m(x) a^\dagger(x) |\emptyset\rangle_a |0\rangle_c |m\rangle_b + \sum_m C_m |\emptyset\rangle_a |1\rangle_c |\tilde{m}(1)\rangle_b. \tag{6}$$

Here, $|\emptyset\rangle_a$ denotes the vacuum state of the fiber mode, $|n\rangle_c$ and $|m\rangle_b$ represent the number states of the cavity and mechanical mode, respectively. $|\tilde{m}(n)\rangle_b = \exp[ng(b^\dagger - b)/\omega_M]|m\rangle_b$ is the $n$-photon displaced number state of the mechanical oscillator, corresponding to the mechanical state whose equilibrium position is shifted by $-ngx_{ZPF}/\omega_M$, where $x_{ZPF}$ is the zero point fluctuation of the mechanical oscillator. The eigenequation of the optomechanical system reads

$$H_{om}|n\rangle_c |\tilde{m}(n)\rangle_b = (n\omega_c + m\omega_M - n^2 \Delta_{om})|n\rangle_c |\tilde{m}(n)\rangle_b, \quad \Delta_{om} = g^2/\omega_M. \tag{7}$$

The energy level diagram of the optomechanical system is depicted in Fig. 2. By substituting the explicit forms of $H$ and $|\omega_0, m_0\rangle$ into Eq. (5), we obtain the single-photon transport equations

$$\begin{aligned}
&-iv_g \frac{\partial f_m(x)}{\partial x} + (m\omega_M - E) f_m(x) + \sqrt{v_g \kappa_1} \sum_{m'} C_{m'} \delta(x) \, _b\langle m|\tilde{m}'(1)\rangle_b = 0, \\
&C_m(\omega_c + m\omega_M - \Delta_{om} - E) + \sqrt{v_g \kappa_1} \sum_{m'} f_{m'}(0) \, _b\langle \tilde{m}(1)|\tilde{m}'\rangle_b = 0.
\end{aligned} \tag{8}$$

The solution has the form

$$f_m(x) = \begin{cases} \delta_{mm_0} e^{i\omega_m x/v_g}, & x < 0, \\ t_m(\omega_0, m_0) \cdot e^{i\omega_m x/v_g}, & x > 0, \end{cases} \tag{9}$$

where $\omega_m = \omega_0 + (m_0 - m)\omega_M$, and $t_m(\omega_0, m_0)$ is the transmission amplitude. By substituting Eq. (9) into Eq. (8) we derive the transmission amplitude

$$t_m(\omega_0, m_0) = \delta_{mm_0} - i\kappa_1 \sum_{m'} \frac{_b\langle \tilde{m}'(1)|m_0\rangle_b \, _b\langle m|\tilde{m}'(1)\rangle_b}{\Delta_0 - (m' - m_0)\omega_m + \Delta_{om} + i\kappa_1/2}, \quad \Delta_0 = \omega_0 - \omega_c, \tag{10}$$

where $\Delta_0$ is the detuning of the input photon. The transmission probability is given by $T_m = |t_m|^2$.

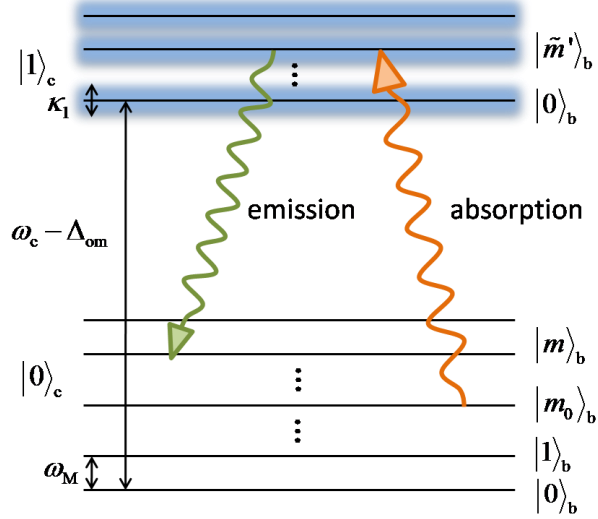

FIG. 2 Energy level diagram of the optomechanical system. The incident photon is absorbed by the optomechanical system and it is later emitted.

The photon transmission is the consequence of the interference between the direct transmission and the cavity re-emission, corresponding to the first and second terms of Eq. (10), respectively. The process of photon absorption and re-emission is described in Fig. 2. When the incident photon enters the cavity, it excites the optomechanical system from initial state $|0\rangle_c |m_0\rangle_b$ to the upper state $|1\rangle_c |\tilde{m}'\rangle_b$ ($m' = 0, 1, 2, ...$). The upper states are broadened by $\kappa_1$ and the resonant transition occurs when $\Delta_0 = -\Delta_{om} + (m' - m_0)\omega_M$. The upper state finally decays into state $|0\rangle_c |m\rangle_b$ ($m = 0, 1, 2, …$), resulting in an emitted photon with a different frequency $\omega_m = \omega_0 + (m_0 - m)\omega_M$.

Next, we depict the spectrum of the transmitted photon with the results given by Eqs. (9) and (10). Generally, a non-monochromatic incident photon can be interpreted as $|1\rangle = \int d\omega \cdot F(\omega) a^\dagger(\omega)|\emptyset\rangle$, where $F(\omega)$ is the spectral density function. The spectrum of the transmitted photon, correspondingly, can be obtained via superposition. In Fig. 3, we consider an incident photon with a Gaussian spectral density $F(\omega) = (2/\pi d^2)^{1/4} \exp\left[-(\omega - \omega_0)^2 / d^2\right]$, and plot the spectrums $S(\Delta\omega)$ of the transmitted photon, where $d$ is spectrum width of the incident photon and $\Delta\omega = \omega - \omega_c$. By changing the coupling rate $g$, the detuning of the incident photon $\Delta_0$, and the initial phonon number $m_0$, the spectra illustrate following points: (i) As expected, more sidebands appear when the coupling strength increases; (ii) The spectra show some dips at resonant transition frequencies $\Delta\omega = -\Delta_{om} + (m' - m_0)\omega_M$ ($m'= 0, 1, 2, ...$). This results from the destructive interference between direct transmission and cavity photon reemission; (iii) As the initial mechanical state changes from $|0\rangle_b$ to $|1\rangle_b$, a blue sideband appears since there is a probability for the photon to increase its energy by absorbing the phonon's energy.

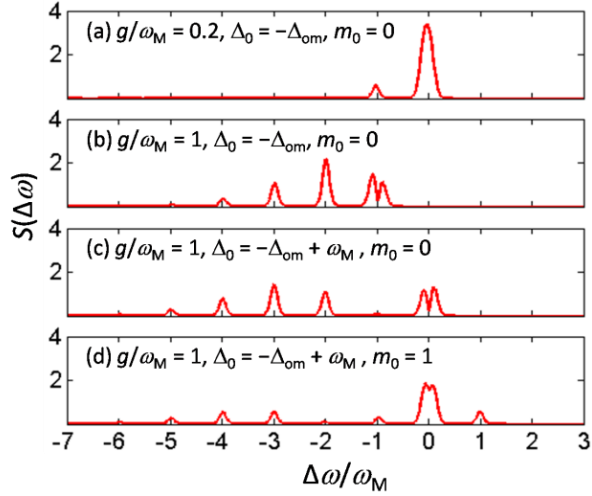

FIG. 3 Transmitted photon spectra $S(\Delta\omega)$ or various coupling rates $g$, the incident photon $\Delta_0$ detuning, and initial phonon numbers $m_0$. Here, $\kappa_1/\omega_M = 0.2$ and $d/\omega_M = 0.2$.

## IV. EFFECTS OF THERMAL NOISE

In the following, we study the influence of the thermal bath on the single-photon transport in the optomechanical system. By neglecting the intrinsic cavity damping $\kappa_0$, the system can be analyzed with the quantum master-equation

$$\dot{\rho}(t) = -i[H,\rho] + \gamma_M(n_{th}+1)D[b]\rho + \gamma_M n_{th} D[b^\dagger]\rho \qquad (11)$$

where $\rho$ denotes the density matrix of the system and $D[b]\rho = b\rho b^\dagger - (b^\dagger b\rho + \rho b^\dagger b)/2$ is the standard dissipator in Lindblad form. To numerically study the problem, we describe the shape of the Gaussian input single-photon pulse by $f(t) \propto \exp\left[-(t-T/2)^2/(T/8)^2\right]$, where $T = 16/d$ is the pulse duration and time $t$ ranges from 0 to $T$. Initially, the mechanical oscillator is assumed at the ground state $|0\rangle_b$. By integrating the master-equation, the transmitted photon spectra within different thermal environments are plotted in Fig. 4. On one hand, as the mean thermal excitation number $n_{th}$ increases, a blue sideband emerges in the transmitted photon spectrum. This is because the mechanical oscillator obtains some excited-state populations through the environment thermal heating. Consequently, the photon can increase its energy by absorbing the energy of the phonon. On the other hand, the resonance dip gradually vanishes when $n_{th}$ increases. This is the result of thermal decoherence which leads to the dephasing between the incident and re-emitted field.

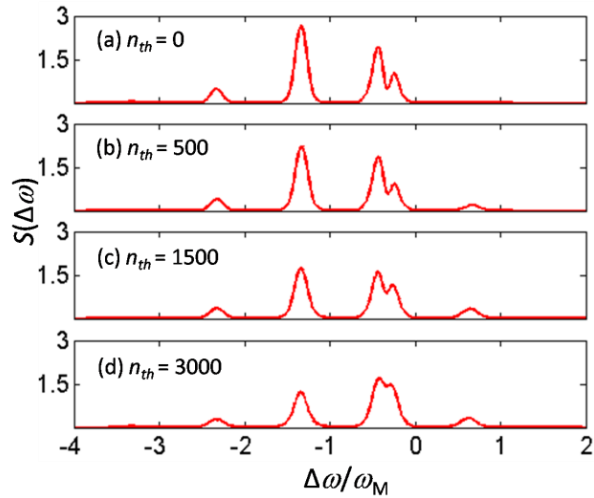

FIG. 4 Transmitted photon spectra $S(\Delta\omega)$ for different mean thermal excitation phonon number $n_{th}$. Parameters are $m_0 = 0$, $g/\omega_M = 0.6$, $\gamma_M/\omega_M = 10^{-5}$, $\kappa_1/\omega_M = 0.2$, $\Delta_0 = -\Delta_{om}$, and $d/\omega_M = 0.2$.

## V. GENERATION OF MECHANICAL NOON STATES

As a potential application of the single-photon optomechanical interaction studied above, we propose a scheme to generate mechanical NOON states $|N::0\rangle_{1,2} = (|N\rangle_1|0\rangle_2 + |0\rangle_1|N\rangle_2)/\sqrt{2}$, which contain $N$ phonons in an equal superposition of all being in one of two remote mechanical oscillators 1 and 2. These states are maximally entangled states and are close to the "cat states" originally envisioned by Schrödinger [34]. Generation of such states in macroscopic oscillators is required for the experimental studies of fundamental physics, such as the effects of decoherence on many-particle entanglement [35, 36] and the quantum-classical crossover [37].

The setup we consider here is based on two identical optomechanical systems, driven by a single photon source. The mechanical entanglement is created by combining the two output fields [38, 39]. Each of the optomechanical systems is coupled to an optical fiber, as shown in Fig. 4. To generate mechanical NOON states, the two mechanical oscillators are both pre-cooled to the ground states. A single photon input from the left has equal probability to transport through either of the two optomechanical systems due to the first fiber coupler. The transmitted photon then reaches the second fiber coupler and is finally detected by either of the photon detectors. Let the energy of the input photon be $\omega_0$, the detection of an output photon with energy $\omega_0 - N\omega_M$ means that one of the mechanical oscillators has been excited to the $N$-phonon number state. As the photon path information has been erased, the two mechanical oscillators have now collapsed onto the NOON state which contains $N$ phonons.

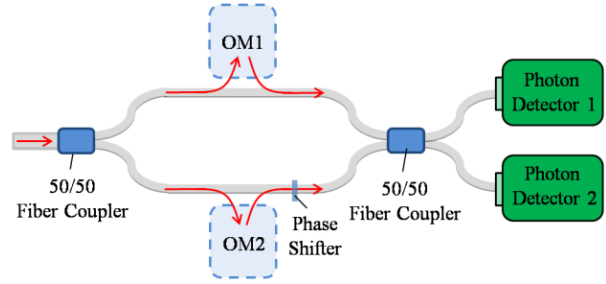

FIG. 4. Setup for generating mechanical NOON states.

Next, we study the probability $P(N)$ of generating the mechanical NOON state $|N::0\rangle_{1,2}$. To obtain the large generation probabilities for different NOON states, the input photon detuning $\Delta_0$ should be specifically chosen. For example, we find that large probabilities for NOON states $N = 1$ and $N = 5$ can be derived when $\Delta_0 = -\Delta_{om}$ and $\Delta_0 = -\Delta_{om} + \omega_M$, respectively. Under these conditions, the dependence of $P(1)$ and $P(5)$ on both $g/\omega_M$ and $\kappa_1/\omega_M$ are plotted in Fig. 5. Here, the input photon is Gaussian with a spectrum width $d = 0.2\omega_M$. It can be seen that $P(1)$ reaches 0.64 when $g/\omega_M = 0.7$, $\kappa_1/\omega_M = 0.6$, and $P(5) = 0.15$ is accessible as $g/\omega_M = 1.4$, $\kappa_1/\omega_M = 0.5$.

Now we turn to investigate the fidelity of the mechanical NOON states. In the scheme, the fidelity of generating state $|N::0\rangle_{1,2}$ is defined as $F_N = \left(\mathrm{Tr}\left(\sqrt{\sqrt{\rho_N^f} \rho_N^{ideal} \sqrt{\rho_N^f}}\right)\right)^2$, where $\rho_N^{\text{ideal}} = |N::0\rangle_{1,2}{}_{1,2}\langle N::0|$ is the ideal mechanical density operator and $\rho_N^{\text{f}}$ is the density operator of the oscillators when a $N$th red sideband photon is detected. Using the master-equation simulations we described in the section above, we derive the fidelities with different parameters. For example, when $\{\omega_m, \gamma_M\} = \{100 \text{ MHz}, 1 \text{ kHz}\}$ and the environment temperature is around 0.2 K, we find that $F_1$ reaches 0.94 as $\{g, \kappa_1\} = \{70 \text{ MHz}, 60 \text{ MHz}\}$, and $F_5$ is 0.82 when $\{g, \kappa_1\} = \{140 \text{ MHz}, 50 \text{ MHz}\}$.

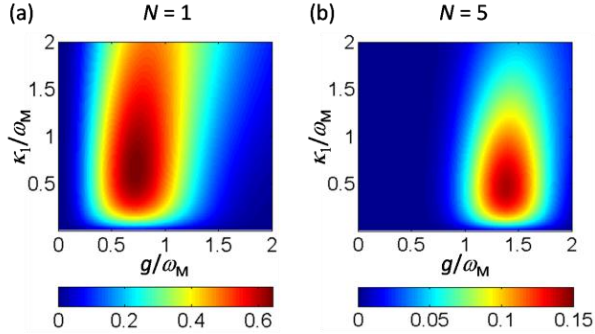

FIG. 5 Probabilities of generating the mechanical NOON states. The incident photon is nearly monochromatic with spectrum width $d/\omega_M = 0.2$. Other Parameters are (a) $N = 1$, $\Delta_0 = -\Delta_{\text{om}}$. (b) $N = 5$, $\Delta_0 = -\Delta_{\text{om}} + \omega_M$.

## VI. CONCLUSION

In summary, we have studied the properties of single-photon transport in a single-mode fiber coupled to an optomechanical system in the single-photon strong-coupling regime. Through a real space approach, the single-photon transmission has been analytically solved. The results explicitly reveal the physical processes of photon absorption and re-emission. The effects of the thermal bath on the single-photon transport are also studied. The output photon spectrum in thermal environment clearly reflects thermal excitation and thermal decoherence. The study provides a detailed understanding of optomechanical interaction and sets a theoretical foundation for single-photon manipulations in optomechanics, e.g., generation of mechanical NOON states.

**ACKNOWLEDGMENTS**

This work was supported by NSFC (Nos. 11222440, 11004003, and 11121091), 973 program (No. 2013CB328704), and RFDPH (No. 20090001120004). X.X.R., H.K.L., and M.Y.Y. were supported by the National Fund for Fostering Talents of Basic Science (Nos. J1030310 and J1103205), the Undergraduate Research Fund of Education Foundation of Peking University, and.the Chun-Tsung Scholar Fund for Undergraduate Research of Peking University. X.X.R. and H.K.L. contributed equally to this work.